\documentclass[12pt]{article}

\usepackage{graphicx}
\usepackage{scicite}
\usepackage{float}
\usepackage{enumitem}
\usepackage{gensymb}
\usepackage{amsmath}
\usepackage{mathtools}

\usepackage{times}

\title{Tracking Janus microswimmers in 3D with Machine Learning} 
\author
{Maximilian R. Bailey,$^{\ast}$\textit{$^{a}$} Fabio Grillo\textit{$^{a}$}, and L. Isa$^{\ast}$\textit{$^{a}$}\\
\normalsize{$^{a}$Laboratory for Soft Materials and Interfaces, Department of Materials, ETH Z{\"u}rich,} \\ \normalsize{Vladimir-Prelog-Weg 5, 8093 Z{\"u}rich, Switzerland}\\
\normalsize{$^\ast$To whom correspondence should be addressed:} \\
\normalsize{E-mail: maximilian.bailey@mat.ethz.ch;}\\ 
\normalsize{E-mail: lucio.isa@mat.ethz.ch}
}

\date{}

\begin{document} 


\baselineskip24pt


\maketitle

% Keywords: Please provide a minimum of three and a maximum of seven keywords, separated by commas

% Abstract should be written in the present tense and impersonal style (i.e., avoid we), and be at most 200 words long
\begin{abstract}

Advancements in artificial active matter heavily rely on our ability to characterise their motion. Yet, the most widely used tool to analyse the latter is standard wide-field microscopy, which is largely limited to the study of two-dimensional motion. In contrast, real-world applications often require the navigation of complex three-dimensional environments. Here, we present a Machine Learning (ML) approach to track Janus microswimmers in three dimensions, using Z-stacks as labelled training data. We demonstrate several examples of ML algorithms using freely available and well-documented software, and find that an ensemble decision tree-based model (Extremely Randomised Decision Trees) performs the best at tracking the particles over a volume spanning a depth of more than 40 $\mu$m. With this model, we are able to localise Janus particles with a significant optical asymmetry from standard wide-field microscopy images, bypassing the need for specialised equipment and expertise such as that required for digital holographic microscopy. We expect that ML algorithms will become increasingly prevalent by necessity in the study of active matter systems, and encourage experimentalists to take advantage of this powerful tool to address the various challenges within the field.

\end{abstract}

\newpage
\vspace{1cm}

\section{Introduction}

Inspired by the collective phenomena that emerge in biological systems across a range of length scales, active matter systems – agents that transduce freely available energy into directed motion \cite{Ramaswamy2010} – have been the subject of extensive research in soft matter and statistical physics. In particular, the behaviour of active materials at the microscale is of interest due to the interplay between thermal fluctuations and the driving force of motion. This driving force makes active matter systems intrinsically out-of-equilibrium, and endows them with promising properties for applications, ranging from smart drug delivery \cite{Paula2021} to water remediation \cite{Wang2019}. 

Perhaps the most popular model system to study the dynamics of active matter are Janus microswimmers \cite{Zhang2021}, which are also arguably the simplest class of synthetic active agents. These (typically micron-sized) particles possess patches with different chemical or physical properties (e.g. catalytic activity) that can generate asymmetric gradients around the particle under certain experimental conditions, leading to motion by self-phoresis \cite{Golestanian2005,Popescu2010,Dey2016,Bailey2021b}. The dynamics of these Janus microswimmers are typically confined to 2D because of their density and interactions with the substrate, whether they move by self-diffusiophoresis \cite{Uspal2015} or self-dielectrophoresis \cite{Gangwal2008}. However, from an applications perspective, the ability to navigate in 3D is highly desirable \cite{Campbell2013}. As such, there is a growing interest in the synthesis of microswimmers that display 3D motion and in developing greater understanding of their swimming behaviour \cite{Campbell2017,Singh2018e,Yasa2018,Lee2019,Brooks2018}.  

Unfortunately, tracking the motion of microswimmers in the third dimension presents new experimental challenges that must be overcome. Conventional light-microscopy techniques are ill-suited to track the diffusion of micron-sized objects in three dimensions, especially when the particles exhibit enhanced mobility. A widespread approach to studying (fluorescent) colloids in 3D is confocal microscopy \cite{Prasad2007}. This involves scanning the volume of interest via a series of Z-stacks, using a pinhole to exclude signal outside the imaged Z-slice. The resultant intensity distributions of the fluorescent particles are then used to obtain a 3D reconstruction from which the particle centres can be identified. This methodology, although accurate, is limited in temporal resolution by the time taken between subsequent Z stacks \cite{Saglimbeni2014}. In the case of active Janus colloids, which can move several body-lengths per second, this leads to a smearing of the traced particle and thus difficulties in accurate centre finding. Furthermore, the surface patches of Janus microswimmers, typically realized with thin metal films, leads to an asymmetry in its optical properties, hindering the proper reconstruction of the intensity distribution and thus jeopardising the tracking.   

To overcome these issues, Campbell et al. used the method previously proposed by Spiedel et al. for passive particles to track the vertical position of fluorescent gravitactic Janus microswimmers from 2D microscopy images using the outermost radius of the concentric rings of the fluorescent particles as they swim out of focus \cite{Speidel2003,Campbell2016}. Unlike confocal microscopy, this method does not require a rigorous model and knowledge of the optical properties of the microscope. A single Janus particle was fixated in gellan gum, from which the “bright ring radius” of the fluorescent particle was extracted at different heights, to which a calibration curve was fitted using a cubic polynomial. The heights of the Janus microswimmers were then predicted from the evolution of the bright ring radius of the fluorescent Janus microswimmers taken from a single 2D plane. The described method effectively extracted 3D trajectories from 2D videos using a fluorescent microscope and a basic piezo Z-stage, allowing standard microscopy image acquisition conditions. However, it is noteworthy that the radius of the bright rings can be on the order of hundreds of pixels, and as the rings cannot overlap, this limits 3D tracking to very dilute suspensions. Zhang et al. proposed a similar approach to track thermally-diffusing non-fluorescent particles by radially projecting the diffraction pattern of a particle and comparing it to a reference model obtained with a Z-stack \cite{Zhang2008}. Least-squares fitting of the radial profile is then used to determine the Z-height, which minimises the difference between the radial projection of the imaged particle and the reference model. This method is highly sensitive to asymmetries in the particle, and therefore is not viable for a freely-rotating Janus particle. Likewise, the detailed methodology presented by Kovari et al., solving for the vertical position by minimising the difference between the interpolated, continuous look up table (LUT) of radial profiles taken with Z-stacks and the radial profile of the imaged particle, relies on the Mie scattering of symmetric particles \cite{Kovari2019}, and therefore also cannot be applied effectively to microswimmers with significant optical asymmetries. 

To track non-fluorescent particles in 3D, digital holographic microscopy has emerged as a powerful technique capable of digitally reconstructing 3D images with holograms obtained from the interference pattern between the sample and a reference laser beam \cite{Saglimbeni2014}. The methodology has been demonstrated to track the motion of photo-gravitactic microswimmers in 3D \cite{Singh2018}, albeit those that principally swim directly upwards, which minimises the extent of visible cap asymmetry. The optical asymmetry of a Janus particle, and the particle-wise variations present in their surface patches, would otherwise further complicate holographic reconstruction \cite{Ketzetzi2020a}. Midtvedt et al. used Machine Learning (ML) U-Nets, a convolutional encoder-decoder neural network architecture, to interpret in-line holographic data obtained with localization accuracies comparable in performance to off-axis holographic microscopy \cite{Midtvedt2021}. They were thus able to reduce the computational cost of holographic microscopy-based 3D tracking. Their DeepTrack 2.0 software provides a fascinating test case for the applicability of ML to the 3D (and 2D) tracking of particles. Despite this potential decrease in computational cost, holographic microscopy still requires specialised imaging configurations, limiting the general accessibility of this technique. Furthermore, the U-Net is trained on simulated Mie scattering spectra of spherical particles, which does not account for the asymmetries present in Janus particles, which can have implications for accuracy as discussed above. Therefore, a general approach enabling the 3D tracking of spherical non-fluorescent microswimmers with asymmetries in their optical properties is still lacking.

For example, photocatalytic TiO\textsubscript{2}-SiO\textsubscript{2} Janus microswimmers synthesised in large quantities via Toposelective Nanoparticle Attachment (TNA), were recently shown to possess unusual 3D mobility, characterised by predominantly 2D motion inter-dispersed with ballistic out-of-plane segments \cite{Bailey2021a}. Unsurprisingly, the asymmetric optical properties from the TiO\textsubscript{2} nanoparticle cap introduced significant difficulties in accurate 3D tracking by standard methods. Therefore, a simple approach using the particle's grayscale moment of inertia, which varied as a function of its vertical position, was developed. Similar to the strategy in \cite{Campbell2016}, an LUT of the microswimmers' evolving moment of inertia was extracted from a series of particle Z-stacks, to which a cubic polynomial was fitted (see Supporting Information of \cite{Bailey2021a}). However, the over-reliance on a singular image property averaged over a few particles is undesirable, and motivates a more robust approach to the 3D tracking of non-fluorescent Janus microswimmers with conventional wide-field microscopy techniques.

We thus return to ML, previously discussed in reference to the Deeptrack 2.0 software, as a promising approach to 3D microswimmer tracking. In contrast to conventional statistical approaches that assume an appropriate data model, ML models algorithmically learn the relationship between a target response and its predictors \cite{Breiman2001}. ML can thus be more generally described as a way that machines can learn to perform tasks without specific programming or a set of rules to follow. Inspired by the ability of ML models to detect underlying patterns in data, we here investigate the suitability of traditional ML techniques for 3D tracking from standard 2D brightfield microscopy images. Relevant features are extracted from light-microscopy videos, and the underlying structure characterising their vertical position are extracted using the freely accessible \textit{Scikit-learn} package \cite{Pedregosa2011}. To evaluate our method, we study the classic non-fluorescent Pt-SiO\textsubscript{2} particle system (R = 1.06 $\mu$m) with a clear optical asymmetry, as an example of a challenging but common example of Janus microswimmers whose vertical position would not be easily identifiable with conventional methods (See Figures \ref{fig:Fig1}a,c). We take Z-stacks of the Pt sputter-coated particles freely diffusing in pure water (i.e. passive due to the absence of a fuel), rather than “sticking” them to the glass slide. This likely increases the labelling error of the Z-stacks, but better represents the experimental conditions of the mobile states and the ability of the active colloids to rotate in 3D as they swim \cite{Bailey2021a}. We conclude by studying a different TiO\textsubscript{2}-SiO\textsubscript{2} particle system obtained by TNA with a modified microscope configuration, and demonstrate that traditional ML models can provide a viable and generalisable approach to track non-fluorescent Janus microswimmers moving in 3D. 

\section{Experimental Method}

\subsection{Particle Synthesis}
\subsubsection{Pt-SiO\textsubscript{2} microswimmer synthesis}
Pt-SiO\textsubscript{2} Janus particles were synthesised following well-established protocols \cite{Howse2007}. Briefly, monolayers of SiO\textsubscript{2} microparticles were prepared by spreading a 50 $\mu$L droplet of SiO\textsubscript{2} microparticles (0.5 \% w/w, 2.16 $\mu$m microParticles GmbH) onto a glass-slide, which was pre-treated in a plasma-oven to increase its wettability. A thin, 5 nm Pt film was then sputter-coated onto the monolayer to obtain asymmetrically functionalised Pt-SiO\textsubscript{2} Janus particles. The particles were collected by sonication for 1 minute followed by multiple rounds of centrifugation in water.
\subsubsection{TiO\textsubscript{2}-SiO\textsubscript{2} microswimmer synthesis}
TiO\textsubscript{2}-SiO\textsubscript{2} microswimmers were prepared as previously described \cite{Bailey2021a}. Briefly, Pickering SiO\textsubscript{2}-Wax emulsions were prepared from 250 mg SiO\textsubscript{2} (5 \% w/w, 2.16 $\mu$m microParticles GmbH) suspensions in a 10.8 mg/L didodecyldimethylammonium bromide (DDAB) solution, using a 1:10 molten wax:water volumetric ratio \cite{Perro2009}. The suspension was heated to 75\degree C then stirred for 15 min at 3000 RPM before vigorous mixing at 13500 RPM for 160 s using an IKA T-25 Digital Ultraturrax. After the emulsification step, the Pickering emulsion was immediately placed in an ice bath to rapidly solidify the colloidosomes. The emulsion was then washed in 0.1 M NaCl solution to remove surfactants, before further washing in deionised water. The SiO\textsubscript{2}-Wax colloidosomes were dispersed overnight by gentle agitation in an aqueous solution of a post-modified (poly)pentafluoroacetate (pPFPAC) polymer \cite{Serrano2016}. The pPFPAC-colloidosomes were then washed thoroughly in deionized water before redispersion in a phosphate-buffered saline (PBS) pH 7.0 suspension containing the TiO\textsubscript{2} (P-25 aeroxide) nanoparticles. After gentle mixing overnight, the TiO\textsubscript{2} functionalized colloidosomes were collected by filtration and the wax was removed with chloroform to obtain the final microswimmers.

\subsection{Image Acquisition}
\subsubsection{Pt-SiO\textsubscript{2} Z-stacks}

280 uL dilute solutions of the Pt-SiO\textsubscript{2} in milliQ were pipetted into a flow-through cell (cell 137-QS; Hellma Analytics) with a light path length of 1 mm. Particles were imaged on an inverted microscope (Nikon Eclipse Ti2e) under Köhler illumination with white-light using a 40× objective (CFI S Plan Fluor ELWD 40XC) with adjustable collar (set to 1 mm), and Z-stacks were taken with an exposure time of 30 ms using a Hamamatsu C14440-20UP digital camera. The Z-labels were then adjusted to account for refractive index mismatches between the air objective and aqeuous media as in \cite{Visser1994}.

\subsubsection{TiO\textsubscript{2}-SiO\textsubscript{2} Z-stacks}

280 uL dilute solutions of the TiO\textsubscript{2}-SiO\textsubscript{2} in milliQ were pipetted into a flow-through cell (cell 137-QS; Hellma Analytics) with a light path length of 1 mm. Particles were imaged on an inverted microscope (Nikon Eclipse Ti2e) under Köhler illumination with white light using a 40× objective (CFI S Plan Fluor ELWD 40XC) with adjustable collar (set to 1 mm), and Z-stacks were taken with an exposure time of 30 ms using a Hamamatsu C14440-20UP digital camera. To simulate the conditions of swimming experiments, the particles were also illuminated with UV (340 nm), using a Lumencor SPECTRA X light engine as the excitation source through the objective (epifluorescence). The Z-labels were then adjusted to account for refractive index mismatches between the air objective and aqueous

 media as in \cite{Visser1994}.

\subsubsection{3D motion experiments}

280 uL dilute solutions of the TiO\textsubscript{2}-SiO\textsubscript{2} in fuel-rich aqueous conditions (H\textsubscript{2}O\textsubscript{2}, 3 \% v/v) were pipetted into a flow-through cell (cell 137-QS; Hellma Analytics) with a light path length of 1 mm. The particles were imaged on an inverted microscope (Nikon Eclipse Ti2e) under Köhler illumination with white light using a 40× objective (CFI S Plan Fluor ELWD 40XC) with adjustable collar (set to 1 mm), and videos were taken with an exposure time of 30 ms using a Hamamatsu C14440-20UP digital camera at 10 FPS. To activate the TiO\textsubscript{2} photocatalyst and induce swimming, particles were illuminated with UV (340 nm), using a Lumencor SPECTRA X light engine as the excitation source through the objective (epifluorescence). Particles were imaged 26.6 $\mu$m below their focal plane, to maximise the effective range over which their 3D motion could be tracked. 

\subsection{Image pre-processing and extraction of relevant particle features for model training}

Before training the ML models, it is first necessary to extract out the relevant image features to reduce the computational complexity of the algorithmic learning process, and remove spurious or otherwise non-instructive information (e.g. background noise). Briefly, Z-stacks of the diffusing particles are first taken as described above. The raw image Z-slices contain multiple particles in the field of view, which must first be localised. As our method relies on the interference patterns of the particles, it is necessary that they are relatively well spaced from each other and from the edges, and therefore particles not meeting a threshold separation distance (3 particle diameters) between each other or the edges of the image are removed. A square around each remaining particle centre (mask) is then extracted from the images, and labelled with its corresponding height (Z-slice). The masks are then adjusted to possess the same mean grayscale intensity, before the application of a median filter with a 3x3 kernel size, and then normalisation of the pixel values to take values between 0 and 1. The Z-labels are then adjusted to match a reference image plane, due to difficulty in experimentally ensuring the exact same labels between different Z-stacks. From the adjusted and processed masks, key image features are extracted to reduce the dimensionality of the inputs into the model for training. In this manner, a small vector can be used to represent a much larger particle mask, significantly reducing computational time and sensitivity to noise. These vectors, which are labelled according to the adjusted Z values, are then randomly shuffled and separated into a Training and Test set, before a final feature engineering step to further reduce the dimensions of the input parameter space. As outlined later, the model learns on the training set, and the quality of its predictions are evaluated on the Test set. By preventing the model from learning from the test data, this allows a better evaluation of the generalisability of the predictions.

\subsubsection{Extraction of particle masks from Z-stacks and pre-processing}
Prior to extracting the features from the labelled Z-slices, we first performed a pre-processing step on the raw image data. Particle centres were localised in the field of view using the MATLAB implementation of the Hough circle transform \cite{Yuen1990}, which we found to be the most effective over the largest range of Z-stack slices. Due to the asymmetry of our (non-fluorescent) particles, the underlying assumptions of the more commonly used centroid method do not hold \cite{Crocker1996}. The vertical depth over which centre finding is accurate is highly dependent on the optical configuration of the microscope and should be determined empirically; for the Pt-SiO\textsubscript{2} system, this was a range of approximately 35 $\mu$m, however, in the case of our TiO\textsubscript{2}-SiO\textsubscript{2} microswimmers, we were able to reach up to 40 $\mu$m by adjusting the illumination conditions. A square mask, the size of which is dependent on the microscope resolution and particle dimensions, was then extracted around the particle centres to obtain a labelled set of particle masks. The particle masks were adjusted to all have a mean pixel intensity of 145 units (8-bit grayscale), an edge-preserving median filter transform with a 3x3 kernel size was then applied, and finally the pixel values were re-scaled to 0-1 before exporting the masks for further processing as necessary. 

\begin{figure*}
\centering
  \includegraphics[width=0.9\linewidth]{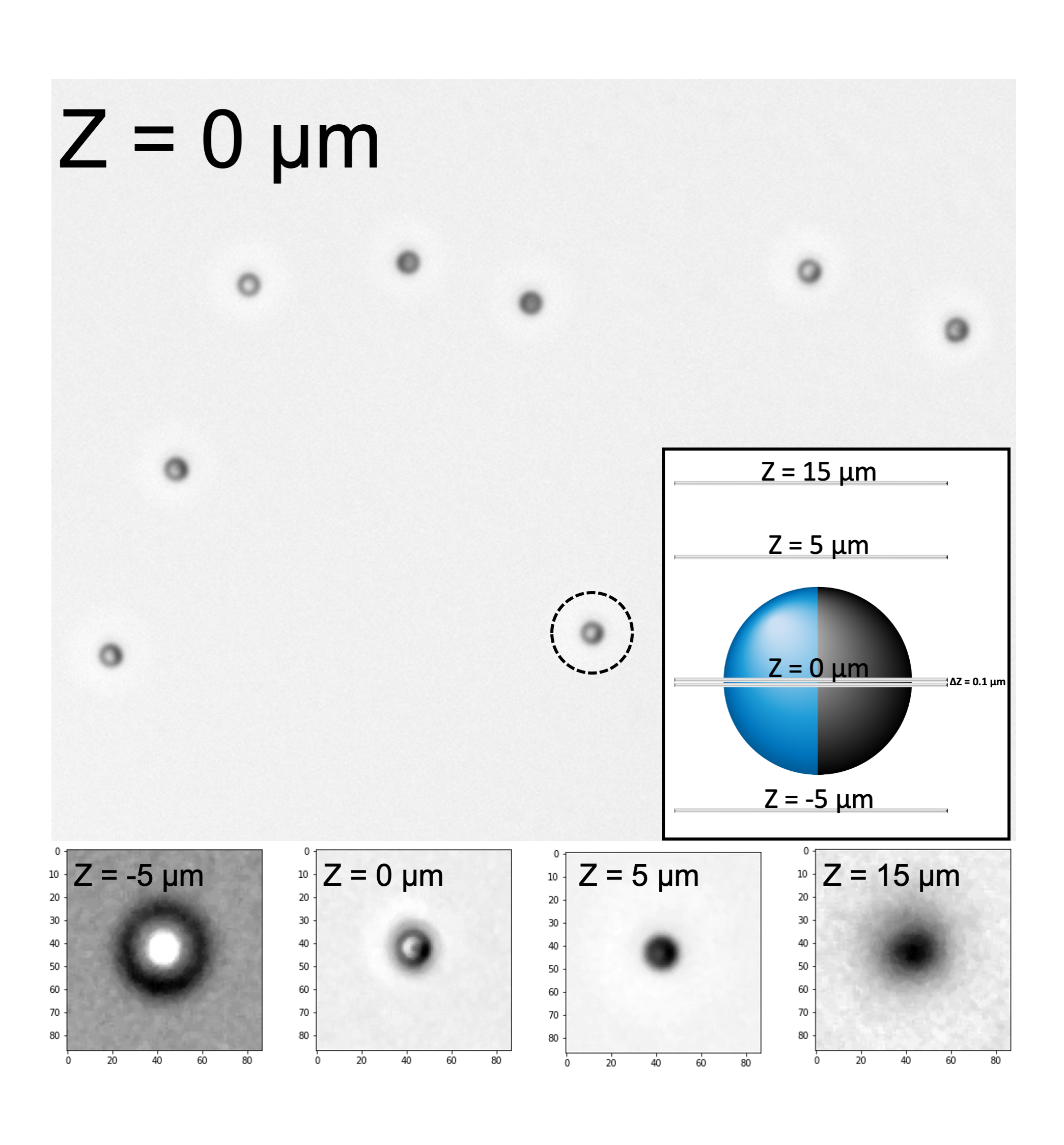}
  \caption{Optical micrograph from a Z-stack of passive (no fuel - diffusing by Brownian motion only) Janus particles to obtain labelled data sets. The optical asymmetry due to the Pt metal film is clearly visible. Snapshots of different Z-slices are shown to demonstrate the changes to the extracted raw particle masks with at different Z values.}
  \label{fig:Fig1}
\end{figure*}

As the model was trained from experimental Z-stacks (rather than simulated data), it was important to ensure that the particle masks are labelled as consistently as possible. We observed that the diffusing particles occupy different vertical positions above the objective, based on their appearances in the microscopy images. Furthermore, it is difficult to reproducibly assign the same focal point ($Z = 0 \mu$m, see Figure \ref{fig:Fig1})   of particles at different regions of the glass cell across Z-stacks taken. By visual inspection of many particle Z-stacks, we noted that each particle defocuses into a black sphere approximately 5 $\mu$m above the slice where it is in-focus (See Figure \ref{fig:Fig1}). This provides us with a reference height that can be compared between the different Z-stacks acquired, to ensure consistent labelling between the Z-stacks and thus the data observations. To use this feature at $Z = 5\mu$m as the reference plane, we examine 5 different Z-stacks, and select for each a representative particle mask where the particle first appears as black sphere. We then average these 5 different particle masks to obtain a reference, $Z = 5\mu$m image. We take the radial profile of this created reference mask, which we defined as the average pixel value of the region spanned by a circle with its origin at the centre of the reference image, for different values of the circle's radius (from 1 pixel to 40 pixels, inclusive). We then scan through the Z-stacks, and for each Z-stack we find the 10 Z-slices that were the most similar to the created, reference image. This was achieved by finding the 10 particle masks in each Z-stack which have the lowest squared difference between their radial profile and that of the reference, radial profile. Of these 10 Z-slices, we select the radial profile from the particle with the lowest Z value, and re-scale all values in the Z-stack such that the selected particle mask is now labelled as $Z = 5 \mu$m. We nevertheless note that this could lead to certain mislabelling of data points, and therefore increase the model error determined during its validation.

\subsection{Feature Extraction and Feature Engineering}
The ML models discussed here are trained on features taken from the 2D particle masks, extracted and labelled as described above, to determine the vertical (Z) position of a microswimmer from the image-based inputs provided. The processed particle masks are 111x111 pixels in dimension (grayscale, 0-1 normalised values), and therefore flattening the image directly as an input vector would create a large feature space with 12321 dimensions. This would significantly slow down training, introduce large amounts of noise from unnecessary image information, and increase the risk of overfitting and thus poor model generalisability. Therefore, it is first necessary to reduce the dimensionality of the input data to improve the training step. This can be achieved by computing various metrics from the particle masks which capture their key features. We identify the following image properties which are initially extracted from the particle masks.

\begin{enumerate}[itemsep=2pt,parsep=0.5pt]
    \item The radial profile of the particle mask
    \item The difference in the maximum and minimum values of the radial and the horizontal profiles of the particle masks
    \item The number of local minima and maxima in the horizontal profile of the particle masks
    \item The image moment of inertia from the particle mask
\end{enumerate}

The radial profile of a particle mask from its centre (as described above), can be used to detect the presence of the interference pattern observed in the images \cite{Kovari2019}. To evaluate the radial profile (40 pixels), we took the mean of the pixel intensities of the entire circle spanned by the specified radius from the centre, rather than individual concentric rings as often performed in the literature. This smoothed the noise arising from the wide-field optical configuration used, but more importantly dampened the significant asymmetry in the optical properties arising from the Janus structure of the microswimmers. From the radial profile of each Z-slice particle mask, we also extract the difference between the maximum and minimum value of the radial profile as an additional, potentially important feature. We likewise extract the difference between the minimum and maximum value of the horizontal profile of each particle mask, obtained by averaging across all vertical pixels of the particle mask for each pixel along the horizontal axis.

The horizontal profile is noisier than the radial profile, largely due to the lack of self-similarity present from averaging over an increasingly growing circle, but it is also more sensitive to the fringes of the interference pattern. We therefore smooth the horizontal profile before determining the number of local maxima and minima present, using the inbuilt MATLAB functions \textit{islocalmin} and \textit{islocalmax}. The number of minima and maxima is not a continuous relationship, and therefore these features are treated as categorical variables. All other features described and used here are treated as numerical, continuous variables. As they take discrete values, the categorical features (number of minima and maxima), were first treated to be properly included in the trained models (One-Hot-Encoding, OHE). From the distribution of the number of maxima and minima, we find that observations with more maxima and minima than 3 (in both cases) are outliers, and therefore set all values for the maxima and minima $>$ 3 to 3. This left two categorical variables which have values of 0, 1, 2, or 3 (8 total features - each with an associated binary value). 

\begin{figure*}
\centering
  \includegraphics[width=0.9\linewidth]{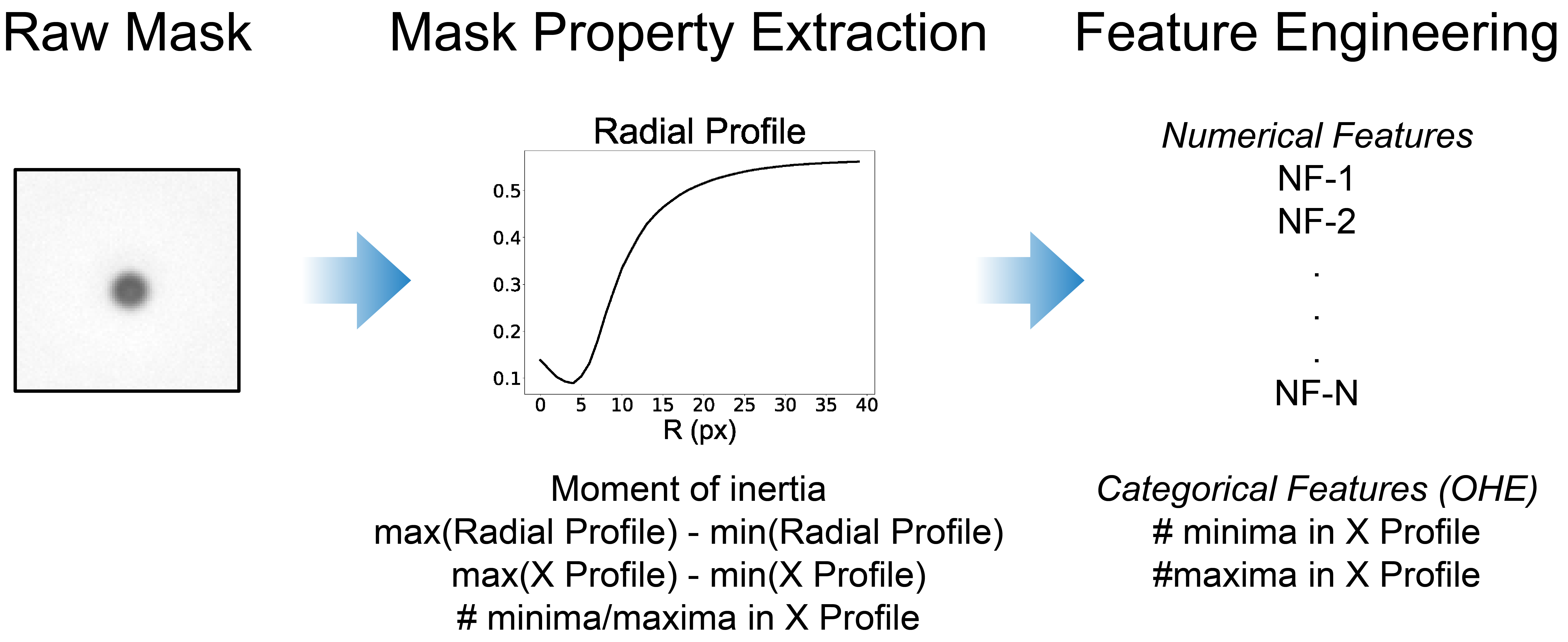}
  \caption{Schematic for extraction of relevant features for model training from raw microscopy videos. Masks around particle centres are first created and processed before the relevant image properties are obtained. After reducing the variable space, the desired features are selected.}
  \label{fig:Fig2}
\end{figure*}

Finally, the particle mask's first image moment \cite{Hu1962}, also referred to as the image's moment of inertia (as it is analogous to the moment of inertia around the image centroid, treating pixel intensities as mass), is calculated for each particle mask \cite{Bailey2021a}. Thus, from the initial 111x111 particle mask with associated Z-label, 51 predictor variables are initially obtained (43 numerical variables and the 8 categorical variables). Next, we perform feature engineering of the numerical variables to further reduce the dimensionality of the feature space and reduce the potential for overfitting (see Figure \ref{fig:Fig2}).

Before performing feature engineering, we first randomly shuffled and split the data into Training and Test sets. Preventing the model from seeing the Test dataset during the training stage ensures that no information from the Test set can influence the outcomes of the model training. This is necessary to test the generalisability of the model predictions. We randomly shuffled all Z slice observations of the particle masks, then split the initial dataset into 80\% for training the models, and 20\% for validating its predictions. We then perform feature engineering on the 43 numerical variables identified previously. From visual inspection, we observe that the distributions for the sequential values of the radial profile in particular appear to show significant self-similarity (see Supporting Information, Figure S4). This can be explained by the method used to extract the radial profiles from the particle masks, as we evaluate the mean pixel intensity of a circle of increasing size centred at the origin. Therefore, sequential values of the radial profile will contain significant amounts of information from the previous values. To treat this collinearity present in the numerical feature space, we use the Python \textit{factor\_analyzer} function to reduce the dimensionality of our feature space (See Figure \ref{fig:Fig2}). The \textit{factor\_analyzer} function implements a varimax rotation and can be used to identify underlying latent variables which capture the largest amount of variance amongst the original feature space. This in turn allows a significant reduction in the number of numerical parameters to be input into the model. The number of extracted features that we use depends on the regression model applied, as we explain in further detail later. 

After extracting the relevant feature vector from our Z-stack images as described above, we fitted the selected model on the Training dataset, or tuned its hyper-parameters to improve performance where closed-form solutions do not exist. The trained model is then used to predict the Z-labels of the hold-out Test set of observations, and these predictions are then compared to the actual labels obtained from the Z-stack. To evaluate model performance on unseen data, we use the normalised error $\epsilon = s(z_{measured}-z_{label})/Z_{total}$ (where $\epsilon$ is the normalised error, $s(z_{measured}-z_{label})$ is the sample standard deviation of the residuals, and $Z_{total}$ is the total valid range of tracking, from \cite{Barnkob2015}). Given the diversity of traditional ML models that can be fitted, we focus here on linear models (Linear and Polynomial regression), and a non-linear ensemble decision tree (Extremely Randomised Decision Trees), to investigate the suitability of ML to track the 3D motion of Janus microswimmers. Further discussions of other models that we investigated (Random Forest, XGBoost, and a Voting Ensemble combining the predictions of the Extremely Randomised Decision Trees and XGBoost models), including a brief investigation into the applicability of convolutional neural nets, can be found in the Supporting Information.

\section{Results}

\subsection{Linear Regression}
We begin our study with the simplest ML model, a standard linear regression using the extracted image features to predict the Z-values labelled from the Z-stack. A closed-form solution for the cost function of a linear regression model exists when the mean-squared-error (MSE) is used, known as the normal equation (See Equation \ref{eqn:normaleqn}). However, it is more computationally efficient to calculate the Moore-Penrose inverse of the vectorized form of the MSE using Singular Value Decomposition (SVD) to obtain the feature weights, which is the default approach used by the \textit{Scikit-learn LinearRegression} class. When there is a large feature space, it is nevertheless advisable to use a general optimization algorithm such as gradient descent-based methods. Linear regression models are highly effective when there is a linear relationship between the predictor variables and the target variable. However, there are a set of assumptions when using linear regression which are often not met in more complex datasets. One of these is the absence of collinearities in the feature space, however this is not the case in e.g. the radial profile of the particle masks as described in the previous section. 

Using the Python \textit{factor\_analyzer} function, we automatically extract 9 underlying variables from an initial parameter space of 43 numerical features, in doing so reducing the extent of the collinearity in the numerical parameter space. However, as we wish to reduce the dimensionality of the linear regression for the subsequent attempt at polynomial linear regression, we take the first 5 variables with the highest explanatory power (with respect to the total variance) of the initial numerical variables. Between them, 99.5\% of the variance in the underlying initial numerical feature space is captured. We finally append the categorical OHE features (number of troughs and peaks) to these 5 features, and run the \textit{Scikit-learn} in-built \textit{LinearRegression} class on the training dataset (80\% of all observations after shuffling, 31834 observations) 

\begin{equation}
\boldsymbol{\hat\theta} = (X^{T}X)^{-1}\boldsymbol{\cdot}(X^{T}Y) \\
\label{eqn:normaleqn}
\end{equation}
\begin{center}
\text{where $Y$ is the vector of observations of the dependent variable, }\\
\text{$X$ is the matrix of independent variables for each observation, } \\
\text{and $\boldsymbol{\hat\theta} = \operatorname*{argmin}_\theta ||Y - X\theta||^{2}$ } \\
\end{center}

\begin{figure*}
\centering
  \includegraphics[width=0.9\linewidth]{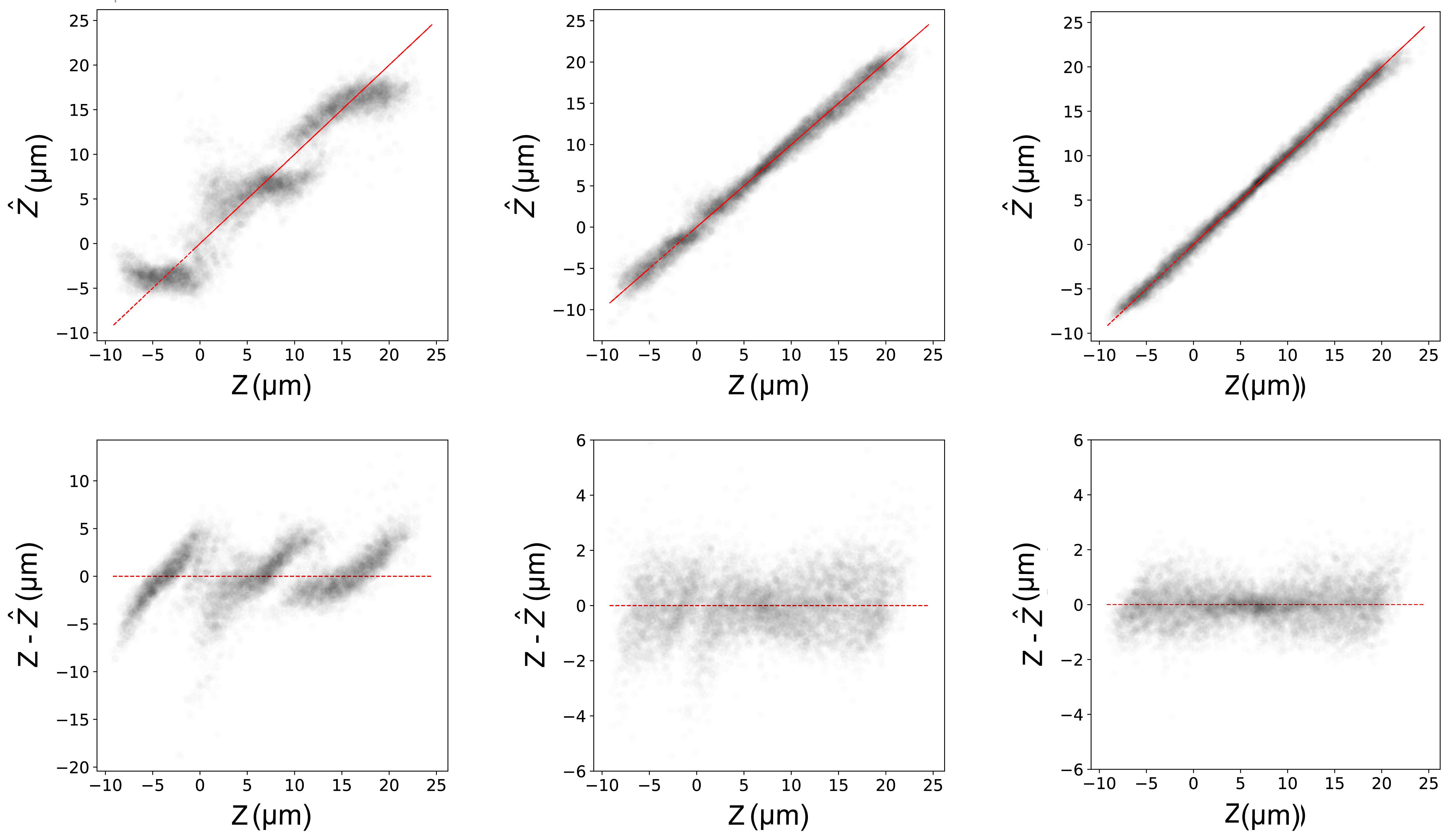}
  \caption{Comparison of the model predictions (top row) and residuals (bottom row) obtained against the ground-truth values from labelled Z-slices, for different regression strategies (39320 total observations). Left: Predictions obtained using linear regression. Middle: Predictions obtained using polynomial regression. Right: Predictions obtained using Extremely Randomised Decision Trees (ERTs). For comparison, the models were trained and tested on the same randomly shuffled datasets.}
  \label{fig:Fig3}
\end{figure*}

From a linear regression model, we obtain a normalised error of 0.080. This translates to an unnormalized value on the order of a particle diameter, which is clearly not acceptable for particle tracking applications. More importantly, we clearly observe that the residuals of the model predictions are non-Gaussian, and that there appears to be an underlying data structure present (see Figures \ref{fig:Fig3}a,b). This demonstrates that a simple linear regression is not sufficient to capture the complexities of the data structure, due to the presence of non-linearities. We therefore move to more complex models, and also do not consider regularization to further improve this simple linear model.

The next ML model that we fit is a polynomial regression model including higher order terms and interactions between the numerical features. In this manner, linear weights can be fitted to non-linear data, and thus a closed-form solution also exists for polynomial regression. The presence of higher order terms and their interactions significantly increases the number of weights to be fitted, and therefore care should be taken in reducing the dimensionality of the feature space and the order of the regression to prevent a combinatorial explosion of parameters (see Equation \ref{eqn:PolyFactors}). The \textit{Scikit-learn PolynomialFeatures} class transforms the input predictor variables to include all higher order terms and interaction terms between each feature. The \textit{LinearRegression} class can then be applied to this transformed data set as before. As the categorical features which are one-hot-encoded are sparse columns with binary values, only the numerical features should be transformed with \textit{PolynomialFeatures}. The categorical OHE features can then be appended to this transformed dataset. From 5 numerical features, transformed to include terms up to the 3rd order, we thus obtain 56 features (and the intercept) to which the 8 categorical features are added. To minimize the possibility of overfitting, we constrain the weights by using Ridge regression. In Ridge regression, a penalty term on the size of the squared weights is added to the cost function to be minimized. This keeps the weights as small as possible, but does not provide feature selection as with LASSO or Elastic-Net regularization techniques. However, it possesses a closed-form solution, making it computationally more efficient and less erratic. Ridge regression can be simply implemented on \textit{Scikit-learn} using the Ridge class, and we use the \textit{GridSearchCV} class to identify the optimal penalty term 

\begin{equation}
\binom{n+d}{d} \quad \text{or} \quad \frac{(n+d)!}{n!\,d!} \\
\label{eqn:PolyFactors}
\end{equation}
\begin{equation*}
\text{where $n$ is the dimension of the original feature space, and $d$ is the highest order of the polynomial} 
\end{equation*}

We find that higher-order polynomial regression provides little to no reduction in training-validation error, and in some cases, can increase it significantly due to over-fitting. We therefore limit our model to a 3rd order polynomial Ridge regression model which we fit to our training data, and then test as previously described. The residuals of the model fitted to the test data are shown in Figure \ref{fig:Fig3}d. We note a significant improvement in the normalised prediction error from the linear regression case to 0.032. Furthermore, we see that most of the systematic periodicity in the residuals present for the standard linear regression has disappeared. Nonetheless, we can still see some structure in the residuals, in particular around $Z = 0 \mu$m where the asymmetry of the Janus particles is likely the most visible.

\subsection{Decision Tree Models}

The inability of linear models to satisfactorily capture the structure of our extracted image properties suggests the use of more complex ML techniques. One powerful and widely used class of models are those based on decision trees. Decision trees make a series of recursive binary splits on the input parameter space, greedily minimising (typically) the Gini impurity (or variance for regression) of the two subsets of data after the split \cite{Altman2017}. Since there is no direct mapping between the predictors and the target variable, Decision Trees, unlike linear regression, can handle non-linearities in datasets. Furthermore, the hierarchical structure of Decision Trees means they can implicitly capture interactions between features \cite{Elith2008}. However, the flexibility of decision trees to model non-linear data is such that they are very sensitive to variations in datasets and prone to overfitting \cite{Myles2004}. To overcome these issues, decision trees are typically integrated into ensembles to average the predictions and thus reduce the model variance at the cost of some bias. So long as the trees are in sufficient number and diversity, even a set of only “weak” learners (low accuracy) can be combined to obtain a “strong” learner (high accuracy) \cite{Schapire2003}. To ensure that the decision trees are sufficiently diverse, a number of strategies can be implemented. Here, we highlight one example of an ensemble decision tree learner – the  Extremely Randomised Decision Trees ensemble model (ERT) \cite{Geurts2006}. For further discussion of the different decision tree ensemble models we investigate, see the Supporting Information.

The ERT model is based on a modification to the well-known random forest (RF) model \cite{TinKamHo1995,Breiman2001RF}, where randomly determined thresholds for each feature at the nodes of a decision tree to split the dataset to improve the scoring metric, rather than determining the optimal thresholds for each feature. Not only does this decrease the variance of the model further compared to a standard RF (at the cost of more model bias), it also significantly reduces the training time for an equivalent RF model, as determining the best thresholds for each feature at each node is one of the most computationally intensive tasks of training. We expect that the high variance present in the input image data makes a model that prioritises variance at the cost of bias more effective at making predictions of the particle's vertical positions. 

We train our ERT ensemble model using the \textit{Scikit-learn ExtraTreesRegressor} class. Unlike the linear regressors, we input all latent features identified during the exploratory factor analysis step (9 factors), which we find reduces the prediction error. With the categorical OHE features, we therefore initially have 17 predictor variables to input into our ERT ML model. To identify potentially non-significant features, we furthermore add a “noise” feature with values drawn from a normal distribution for each observation \cite{Williams2020}. After fitting the ERT, we call the \textit{feature\_importances\_} attribute of the fitted model and find that the observation that the number of peaks and troughs $>= 3$ in the horizontal intensity profile of the particle masks are both less informative to the model than the random noise term. We therefore assign all peaks and troughs $>= 2$ to 2 and re-fit the categorical OHE, reducing the feature space by 2 to 15. Due to the large number of hyperparameters which can be tuned when fitting a ERT (e.g. number of trees, feature sub-set size per node, depth of a tree etc.), we use k-folds (k=10) cross validation using the \textit{GridSearchCV} functionality of \textit{Scikit-learn} to identify the optimal fitting parameters by minimizing the RMSE on the k validation sets.  

After identifying the optimal hyperparameters by K-folds cross validation, we fit an ERT model on the entire training set. We then determine the generalizability of the trained model on the Test dataset, and find an improvement in the model error to 0.021. Furthermore, the structure of the residuals is more homoscedastic than those of the polynomial regression model (see Figures \ref{fig:Fig3}e,f) with only some bias at the extreme values, which was also observed when fitting 3D data with Deeptrack 2.0 \cite{Midtvedt2021}, albeit the explanation for their prediction errors does not apply to our case. The superior performance of the ensemble decision tree models appears to justify their selection, and indicates their suitability for the problem of tracking the 3D motion of Janus microswimmers.

\subsection{Tracking 3D motion}

We investigate the 3D tracking abilities of our ML-based approach and the generalisability of the strategy described here on our photo-responsive TiO\textsubscript{2}-SiO\textsubscript{2} microswimmers. Due to gravity and the existence of well-characterised “sliding states”, the benchmark case of Pt-based microswimmers \cite{Uspal2015,Ketzetzi2020a} typically shows motion constrained to a 2D plane, and we thus expand our tests to a more challenging case of photocatalytic microswimmers \cite{Bailey2021a}. Like the Pt-SiO\textsubscript{2} system we have previously discussed, our TiO\textsubscript{2} microswimmers also possess a challenging (but different) asymmetry in their optical properties. The Janus coating of TiO\textsubscript{2} nanoparticles appears as dark “chunks” under brightfield microscopy (see Supporting Information, Figure S1), allowing us to investigate how well our ML model can handle different types of optical asymmetry. We furthermore use a modified optical configuration with regards to exposure times and brightfield intensity, to check whether our ML workflow can be used across a range of experimental conditions. By adjusting the microscope configuration, we were finally able to track the particle centres over 40 $\mu$m, a wider range than for the Pt-SiO\textsubscript{2} system (35 $\mu$m). 

\begin{figure*}
\centering
  \includegraphics[width=0.9\linewidth]{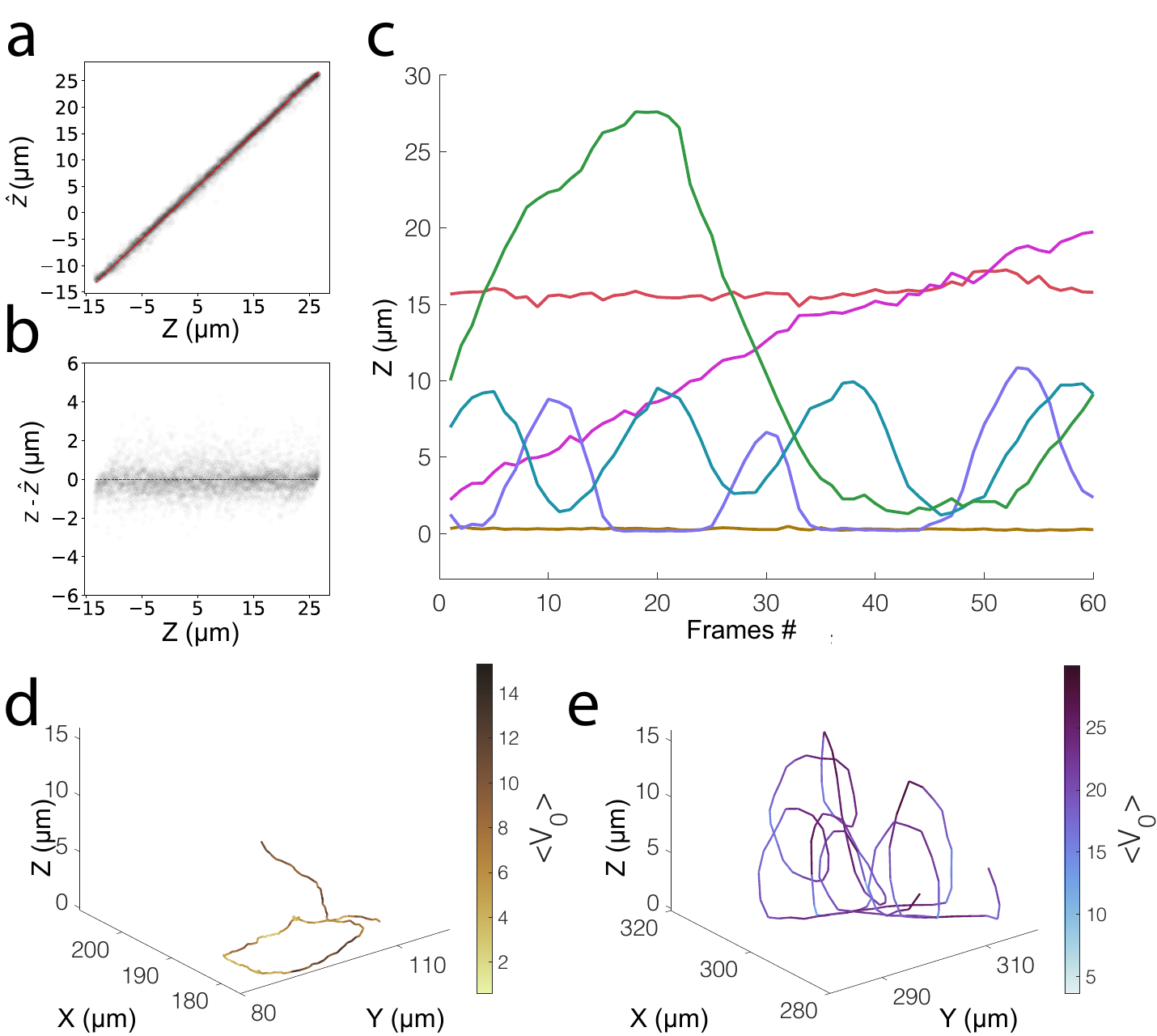}
  \caption{Tracking the 3D motion of SiO\textsubscript{2}-TiO\textsubscript{2} microswimmers using an Extremely Randomised Decision Tree (ERT) model: a) Model predictions vs ground-truth values from labelled Z-slices (26924 total observations, 5385 of which are in the Test set (20\%)). b) Residuals as a function of labelled Z-slices. c) Selected Z positions vs time (different colours denote different microswimmers), comparing the out-of-plane motion of particles for different chiral and non-chiral microswimmers. d) Example trajectory of a microswimmer that remains mostly in-plane. e) Example trajectory of a highly chiral microswimmer that swims out-of-plane}
  \label{fig:Fig4}
\end{figure*}

We extract the same features from the particle masks; however, the exploratory factor analysis stage identified 12 relevant latent factors in this case rather than 9. We attribute this to a range of factors, such as the different optical configuration used, which allows us to scan a wider Z range, as well as the different optical asymmetry of the particles studied. We directly train our ERT on these 12 features extracted using \textit{factor\_analyzer}, and in this manner, we are able to obtain a normalised error of 0.019, with no notable underlying structure in the residuals as shown in Figure \ref{fig:Fig4}a-b.

We thus see that traditional ML models, specifically decision tree-based ensemble models, are an effective approach to track the vertical position of Janus particles and therefore capable of 3D tracking. As a demonstration, we show some sample 3D trajectories that we obtain using our trained ERT model (see Figures \ref{fig:Fig4}c,d,e). Specifically, we show the Z-tracking of a highly chiral “looping” particle, whose motion would present significant difficulty to track using conventional methods. We also show the Z trajectory of a particle that moves in 2D, noting the small length scales over which positional fluctuations occur in the vertical direction. 

\section{Conclusion}

Our findings demonstrate the applicability of simple ML techniques to the 3D tracking of active Janus particles from 2D slices of non-fluorescent wide-field microscopy videos. Rather than simulating the optical configuration of a microscope, our approach allows the training of ML models on Z-stacks taken on a standard light microscope. Although this introduces some uncertainty in the Z-position due to the vertical resolution of the microscope, the performance we achieve using ML models is high, and our method does not require specialised equipment as for holographic microscopy. Furthermore, our approach is robust to Janus microparticles with a high degree of optical asymmetry, which is otherwise a challenge for other 3D tracking techniques \cite{Ketzetzi2020a}. We furthermore expect that higher accuracy could be reported if the models were trained on particles adhered to the glass substrate. However, we wished to replicate true experimental conditions as closely as possible, in particular the rotational diffusion of particles, which exposes different asymmetries to the microscope. Importantly, we find that the trajectories of 2D swimming particles are relatively noise-free in the Z direction, removing most of the spurious displacements which were occasionally present in previous attempts at 3D tracking of Janus microswimmers with a wide-field microscope \cite{Bailey2021a}. We nonetheless stress the importance of checking the model predictions as outlier swimmers will possess noisy trajectories, which should be filtered. The presence of contamination on the glass slide, or even uneven illumination fields, could all contribute to additional sources of noise in the trajectories of swimming particles. It is also critical that the trajectories analysed are limited to the bounds of the optical configuration used (around 40 $\mu$m), and we strongly advise against extrapolating to vertical positions beyond these values. 

We moreover find that traditional ML techniques are effective at tracking the vertical positions of two different Janus particle systems, with videos taken using different optical conditions. We find that decision-tree based ensemble models are the most accurate type of traditional ML model. The extremely randomised decision tree model is the best stand-alone traditional ML technique for predicting the vertical positions of Janus particles, and combining its predictions with those of a Hypothesis Boosting Gradient model in a Voting Regressor may lead to a very slight improvement in overall performance (see Supporting Information, Figure S5). We limit our discussion of DL convolutional neural networks to the SI, but we note that these models can accept particle masks directly, circumventing the time-consuming process of feature extraction. Nevertheless, traditional ML models provide better understanding and control over inputs, and we favour the use of simpler models where possible.

Using the freely-available and well-documented \textit{Scikit-learn} packages, the training of ML models can be performed with a few simple lines of code. Coupled with the extensive learning resources widely available \cite{Geron2019}, this enables a low-barrier extension of existing resources to new problems. We highly encourage experimentalists in the active matter community to explore the fascinating possibilities of ML by developing their own models. If nothing else, it provides an engaging learning exercise in a burgeoning field, but can likely help address experimental challenges in a manner not possible by well-established protocols for particle tracking.

\newpage
% Acknowledgements
\medskip
\textbf{Acknowledgements} \par
The authors thank V. Niggel for the various discussions on image processing and analysis. We also acknowledge the various online learning resources which made this study feasible, as well as the wide-ranging discussions on image-processing and ML to be found on community forums such as Stack Exchange, Stack Overflow, and MathWorks.  

\medskip
\textbf{Author Contribution Statement} \par
Author contributions are defined based on the CRediT (Contributor Roles Taxonomy). Conceptualization: M.R.B., F.G. Formal Analysis: M.R.B. Funding acquisition: L.I. Investigation: M.R.B. Methodology: M.R.B. Software: M.R.B., F.G. Supervision: F.G., L.I. Validation: M.R.B. Visualization: M.R.B., F.G., L.I. Writing - original draft: M.R.B. Writing - review and editing: M.R.B., F.G., L.I.     

\newpage
\bibliographystyle{MSP}
\bibliography{ML_3Dtrack_paper}

\newpage
\setcounter{figure}{0}
\renewcommand{\figurename}{Fig. S}
\renewcommand{\tablename}{Table S}

\section*{Supplementary materials}

\begin{itemize}
    \setlength\itemsep{-1em}
    \item[] Fig. S1: Optical asymmetry of TiO\textsubscript{2} nanoparticles on a SiO\textsubscript{2} microparticle support
    \item[] Fig. S2: Extraction of features from a single particle mask (Pt-SiO\textsubscript{2} Janus particle system)
    \item[] Fig. S3: Distribution of features extracted from all Z-masks (Pt-SiO\textsubscript{2} Janus particle system)
    \item[] Fig. S4: Distribution of radial features extracted from all Z-masks (Pt-SiO\textsubscript{2} Janus particle system)
    \item[] Fig. S5: Performance of different Decision Tree based ensemble methods (Pt-SiO\textsubscript{2} Janus particle system)
    \item[] Text S1: Deep Learning (Convolutional Neural Networks)
    \item[] Fig. S6: Overview of Convolutional Neural Network (CNN) architecture, training, and model performance
\end{itemize}

\newpage

%\paragraph*{Text S1: Pt-film control experiments}
\section*{Traditional Machine Learning models:\\
Pt-SiO\textsubscript{2} Janus particle system}
\subsection*{Image Properties and Feature Extraction}
\begin{figure}[H]
\centering
   \includegraphics[width=0.9\linewidth]{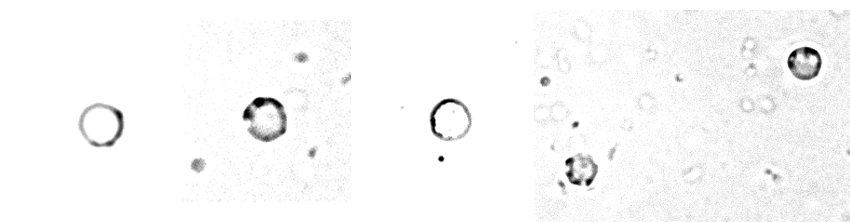}   
  \caption{Optical asymmetries arising from the TiO\textsubscript{2} nanoparticles on a SiO\textsubscript{2} microparticle support. Contrast has been enhanced to highlight the nanoparticles (dark spots)}
   \label{suppfig:1}
\end{figure}

\begin{figure}[H]
\centering
   \includegraphics[width=0.9\linewidth]{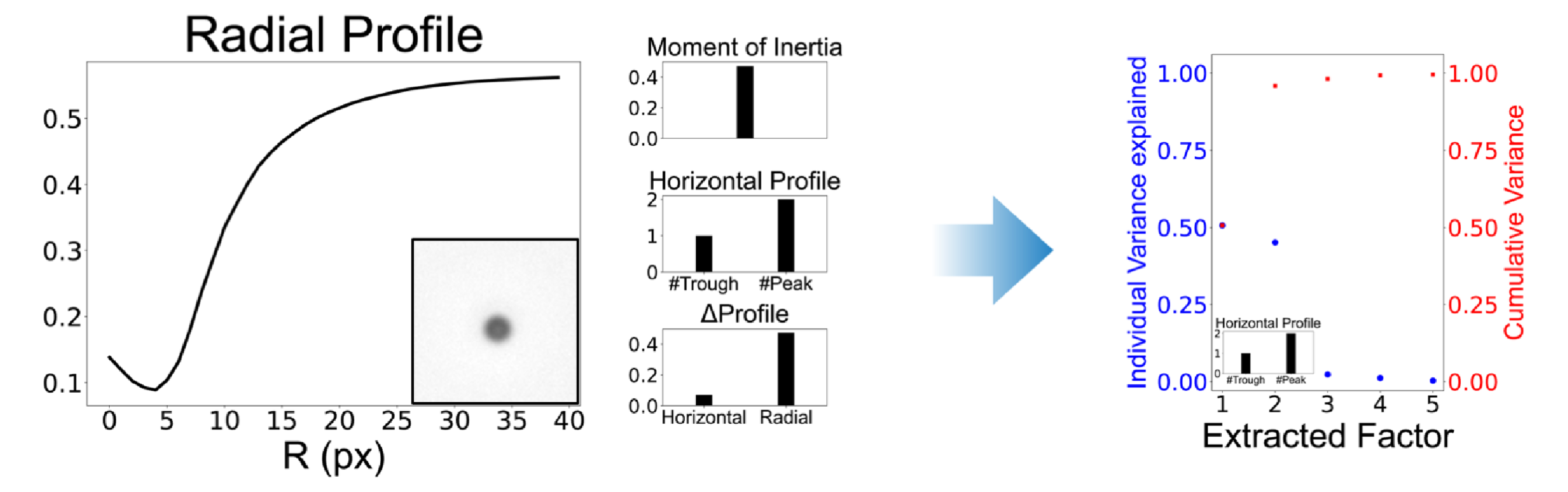}   
  \caption{Extraction of numerical and categorical features from a single particle mask. From the mask, 40 values of the radial profile, the image moment of inertia, the difference between the largest and smallest values in the horizontal and radial profiles, and the number of maxima and minima in the horizontal profile are extracted. From the numerical features (radial profile, moment of inertia, and $\Delta(max-min)$) of the two profiles), exploratory factor analysis determines the 5 underlying factors with the highest explanatory power (99.5\% of total variance) }
   \label{suppfig:2}
\end{figure}

\begin{figure}[H]
\centering
   \includegraphics[width=0.9\linewidth]{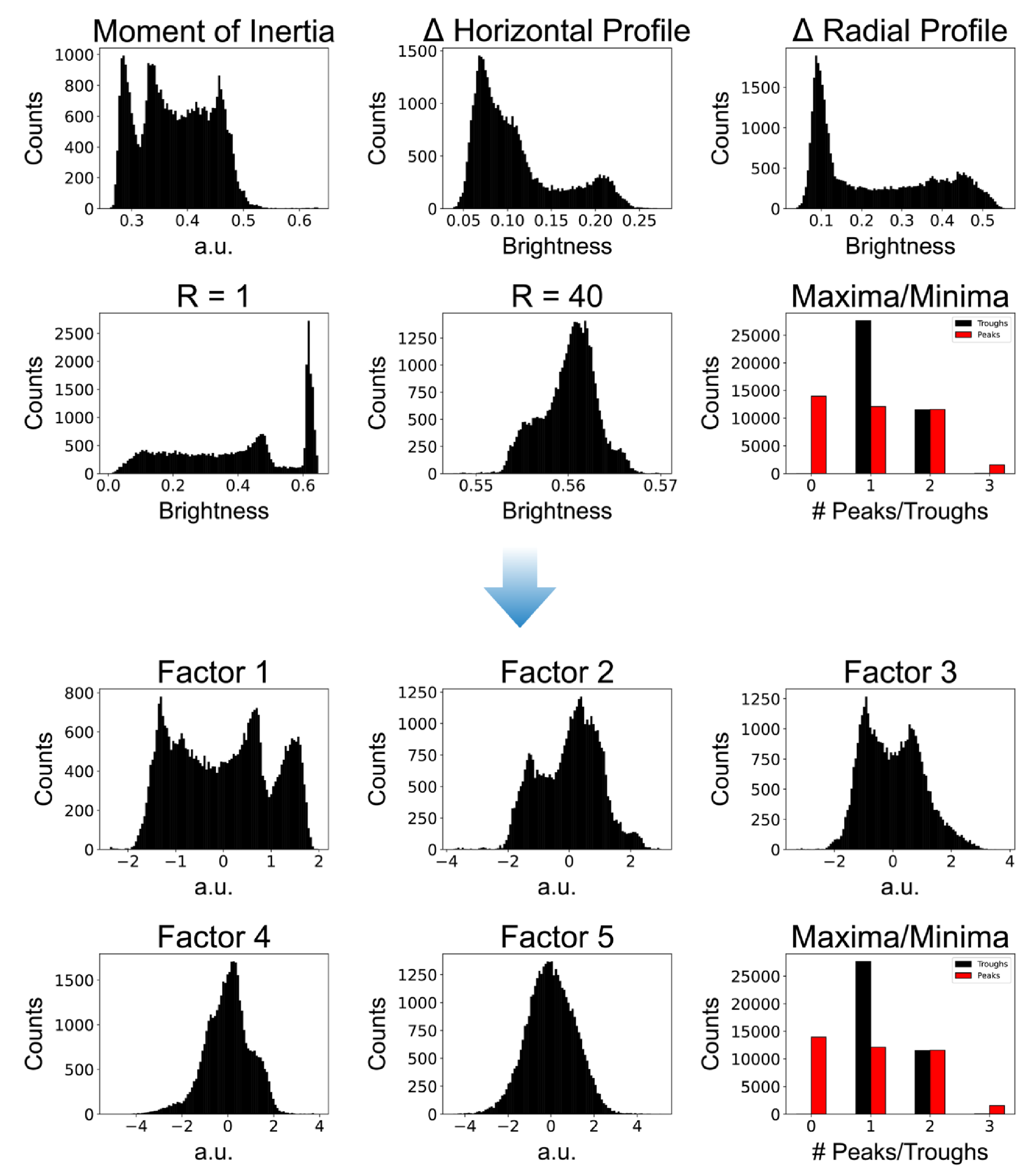}   
  \caption{Extraction of numerical and categorical features from the full set of Z-slices across many particles ($> 100$). The 5 underlying factors with the highest explanatory power are extracted from the numerical features as described in Figure S\ref{suppfig:2}. The distributions across all observations (31834 total Z-slices) for selected features are shown.}
   \label{suppfig:3}
\end{figure}

\begin{figure}[H]
\centering
   \includegraphics[width=0.9\linewidth]{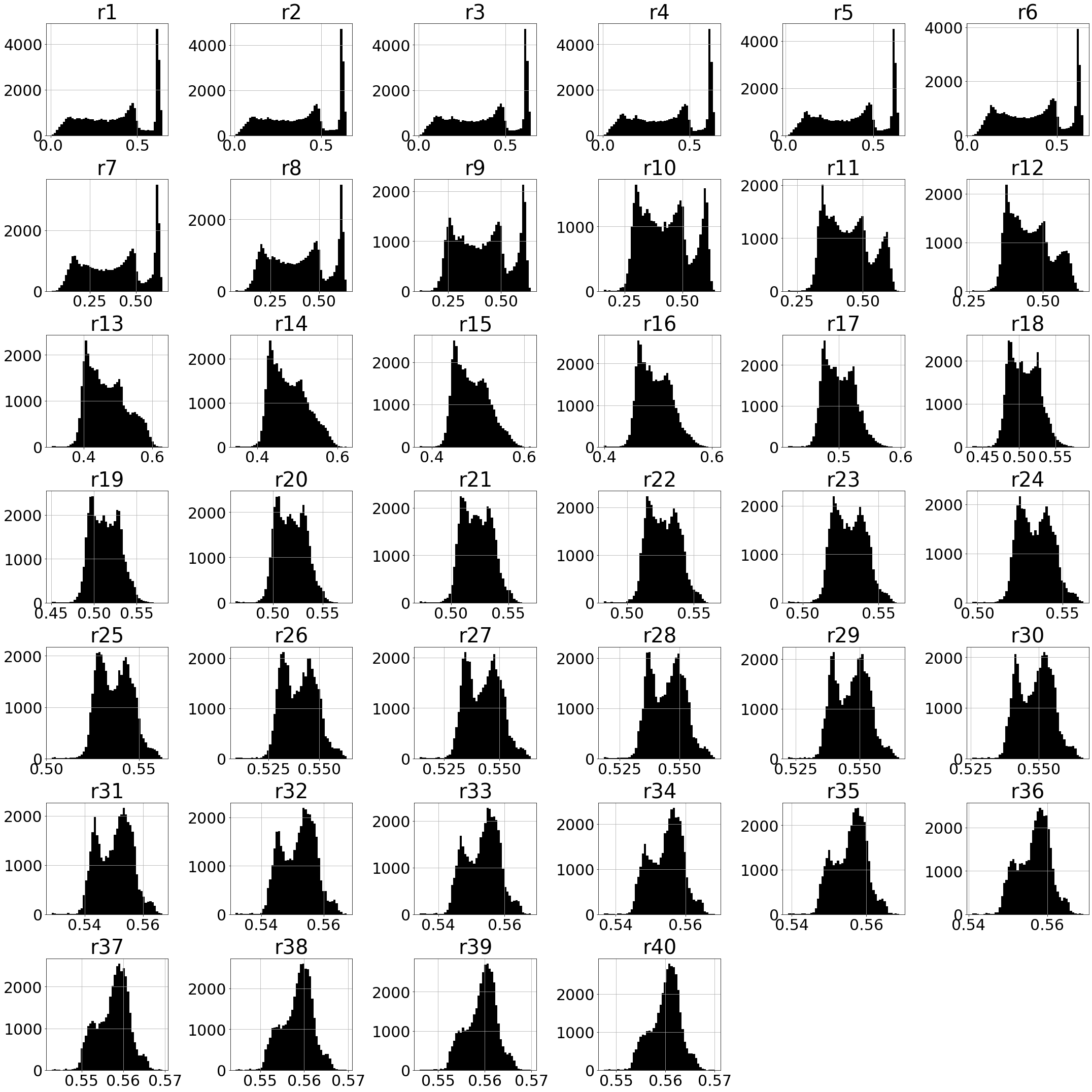}   
  \caption{Extraction of all the radial features from the full set of Z-slices across many particles ($> 100$). The self-similarity between subsequent values of the radial profile can be observed from the distributions}
   \label{suppfig:X}
\end{figure}

\newpage
\subsection*{Performance of Ensemble Tree-based Models}

\begin{figure}[H]
\centering
   \includegraphics[width=0.9\linewidth]{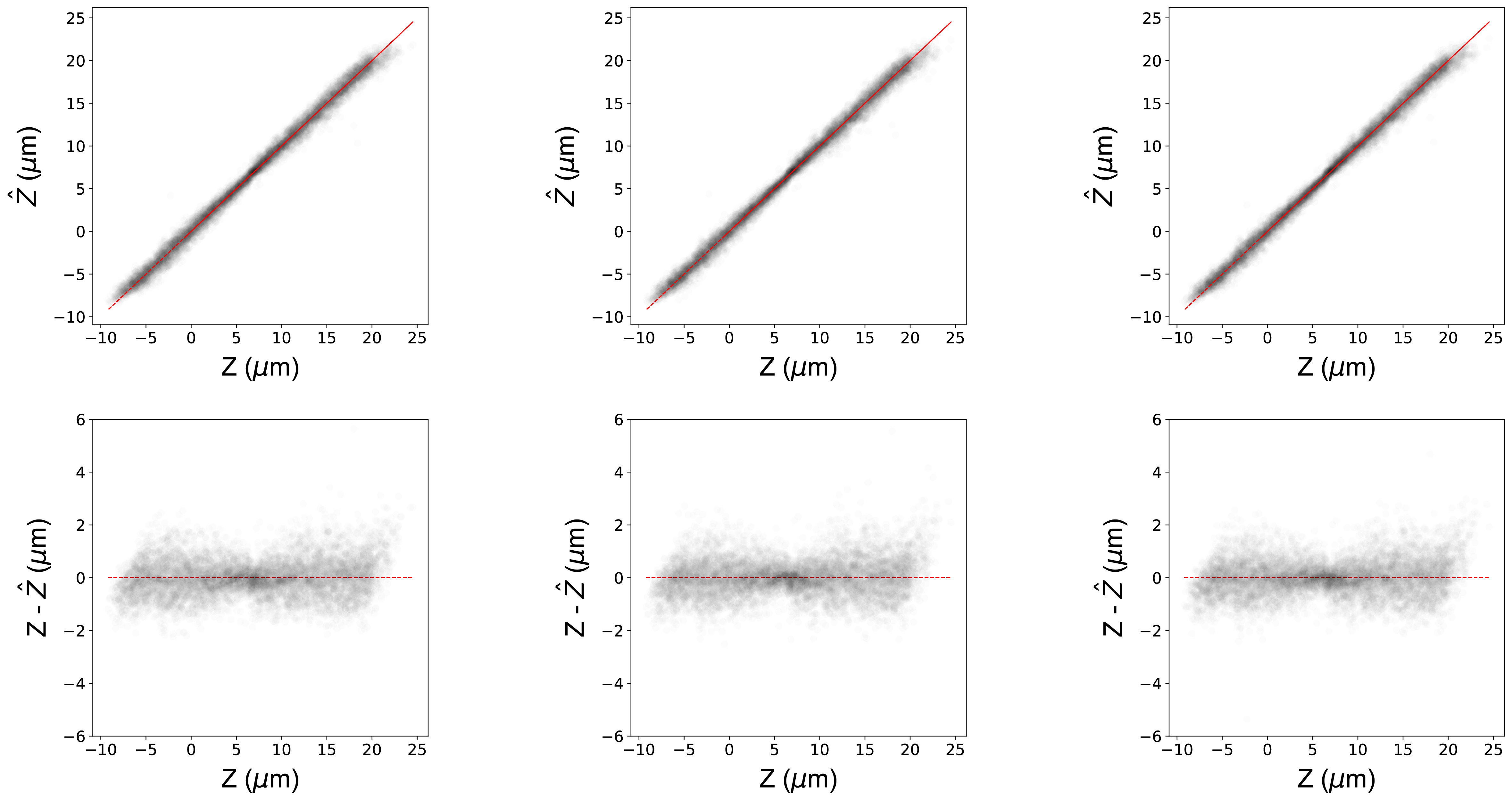}   
  \caption{Comparison of the model predictions (top row) and residuals (bottom row) obtained against the ground-truth values from labelled Z-slices, for different decision tree based ensemble models. Left: Predictions obtained using a 
  Random Forest \cite{TinKamHo1995,Breiman2001}. Middle: Predictions obtained using XGBoost \cite{Chen2016}. Right: Predictions obtained using a voting regressor combining the predictions of the XGBoost model (middle) and Extremely Randomised Trees model discussed in the main text. For comparison, all models were trained and tested on the same randomly shuffled datasets. We find no qualitiative difference in the performance of the respective models (all with normalised errors $\epsilon = s(z_{measured}-z_{label})/Z_{total} = 0.021$ (where $\epsilon$ is the normalised error, $s(z_{measured}-z_{label})$ is the sample standard deviation of the residuals, and $Z_{total}$ is the total valid range of tracking, from \cite{Barnkob2015})). We opt for the Extremeley Randomised Trees in the main text principally by visual inspection of the residuals and the significantly reduced model training time (due to the random selection of thresholds at the decision tree nodes).}
   \label{suppfig:4}
\end{figure}

\newpage
%\paragraph*{Text S1: Pt-film control experiments}
\section*{Text S1: Deep Learning (Convolutional Neural Networks)}% TiO\textsubscript{2}-SiO\textsubscript{2} Janus particle system}

The recent hype surrounding Machine Learning (ML) models is in large part due to the renewed interest in Deep Learning (DL) architectures. The concept of artificial neurons is not new \cite{Mcculloch1943}, but the rise in computing power, demonstrated by Moore’s law, coupled with the accessibility of GPU processors and modifications to training algorithms, has made the training of Neural Networks far more feasible. One of the main advantages of DL models for the 3D tracking of the Janus microswimmers compared to traditional ML models described in this manuscript is that they can directly accept image inputs, and the extensive feature extraction and engineering steps described for the traditional ML models are no longer necessary. This makes the training pipeline of DL models more generalisable to different particle systems and optical configurations compared to traditional ML methods, as we will now demonstrate. 

Amongst standard Deep Learning architectures, Convolutional Networks (CNNs) are the best performers for computer vision tasks such as image classification \cite{Yamashita2018}. The introduction of convolutional and pooling layers into the architecture of neural networks \cite{Fukushima1980,LeCun1998}, was in-part inspired by studies into the neurons of the visual cortex which demonstrated the presence of small, local receptive fields recognising patterns of varying complexity \cite{Hubel1968}. Unlike a dense layer, each neuron in a convolutional layer is connected to a limited subset of pixels in the previous layer. This allows the network to first focus on smaller features before assembling them into higher-level features in subsequent layers. Each convolutional layer will consist of a defined number of filters performing linear operations, outputting different feature maps. Thus, the output of each convolutional layer can be represented as a 3D set of 2D feature map slices representing each linear operation. An activation function, typically the rectified linear activation function \cite{Nair2010}, is then usually applied to the outputs to introduce non-linearities into the transformed feature maps. Pooling layers, which sub-sample the input feature map by aggregating the values and return a feature map of reduced size, are frequently used in CNNs to minimise the memory requirements during training and inference. At the cost of image information, they reduce the number of parameters, computations and memory usage in subsequent layers, and can also introduce a small degree of invariance to different transformations of the image. They are typically used between convolutional layers, particularly in deep CNNs, to prevent an explosion of computational load. The final layers of a CNN are a dense layer with outputs appropriate for the desired regression or classification task.

\begin{figure}[H]
\centering
   \includegraphics[width=0.75\linewidth]{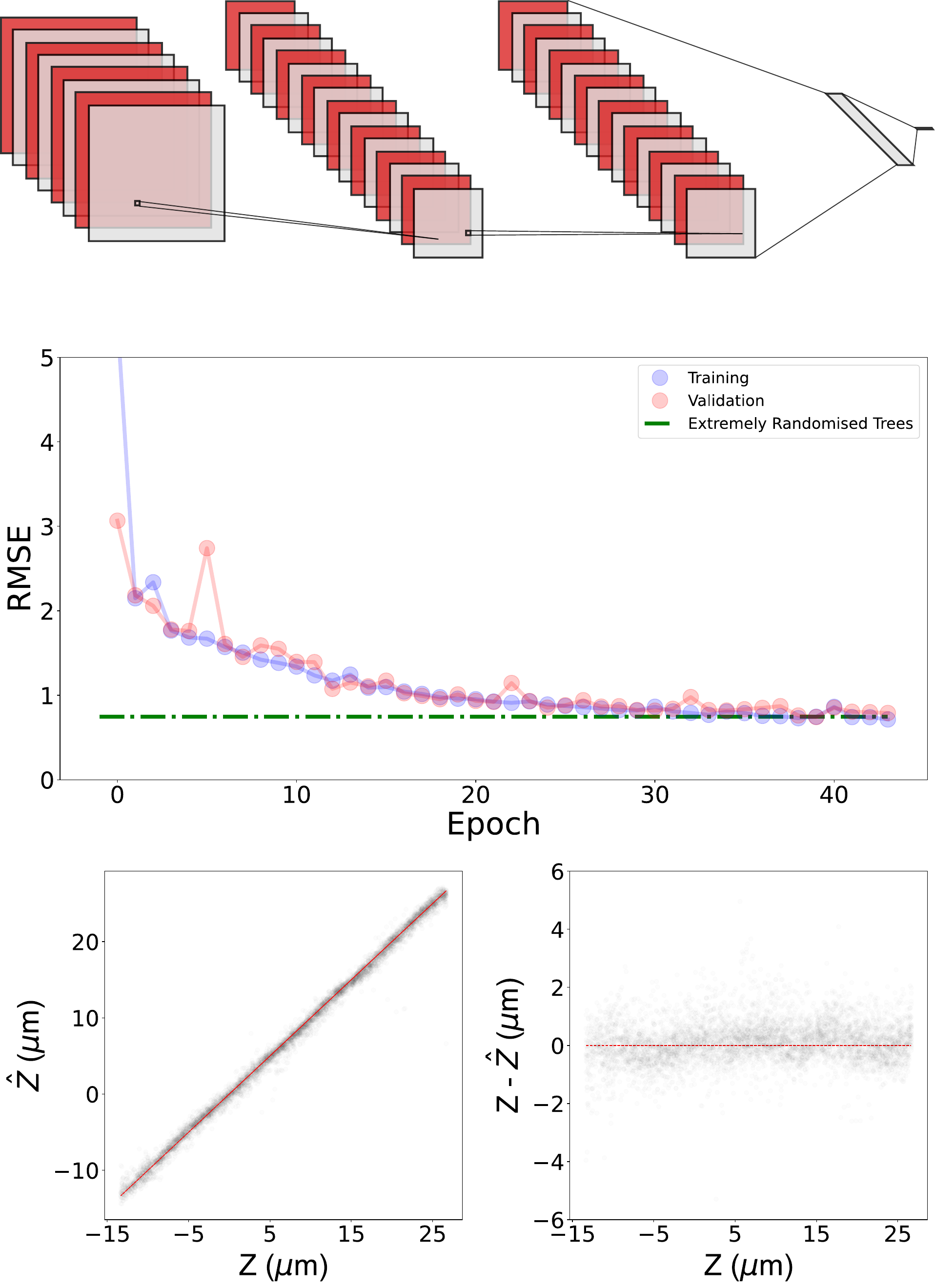}   
  \caption{Top row: Schematic of the architecture of our CNN with 3 Convolutional layers and 2 Dense layers (single output for the Z-label regressed), created using software from \cite{LeNail2019}. Middle row: Training-Validation performance of the CNN with Epochs. The root-mean-squared error (RMSE) saturates at a value similar to that achieved using traditional decision-tree based ensemble ML models. Bottom row: Performance of our trained CNN on the test data-set. We note a slight underlying structure in the residuals (right), which could explain the poor performance on real 3D trajectories despite good performance on the shuffled test data.}
   \label{suppfig:5}
\end{figure}

Given our relatively simple image processing task, we build our CNN sequentially using the Keras backend of the TensorFlow API \cite{chollet2015keras,Abadi2015}, analysing our TiO\textsubscript{2}-SiO\textsubscript{2} Janus particle system. In this manner, we find that the best performance can be achieved with a CNN architecture consisting of 3 convolutional layers and 2 fully connected dense layers (see Figure S \ref{suppfig:5}). The first convolutional layer is constructed with 3x3 kernels outputting 64 feature maps, using a stride of 2. This reduces the dimensionality of the input image (87x87 pixels), allowing an increase in the number of filters in the second convolutional layer (128) without an explosion of parameters. The first two convolutional layers are followed by the ReLU activation function, and no pooling is used. Interestingly, we obtain better performances by introducing a third convolutional layer (128 feature maps) without an activation function. The output of the third convolutional layer is then fully connected to a dense layer, with 64 output neurons and again activated with the ReLU function. The output of this dense layer is finally connected to a dense layer with a single output neuron representing the target Z-value. This output has a linear activation function for regression, which is regularised using L2 (Ridge) penalty terms. The model is trained using Mean Absolute Error (MAE) as the cost function \cite{Midtvedt2021}. Mini-batch gradient descent with early stopping is used to optimise the weights (100 observations), which was the batch size identified as optimal for training and which also easily fit into the memory without significantly extending training times. 

Remarkably, this simple CNN architecture achieves similar results as the Decision tree based ensemble methods that we study, despite the essentially raw particle masks with minimal pre-processing used as inputs. However, in the application to our 3D swimming trajectories, we note significant noise in the axial tracking, specifically during the 2D segments of motion. On closer visual inspection of the model predictions on the test data, we note an underlying structure in the residuals, which has a significant negative impact on our Z-tracking. This highlights the importance of validating the model predictions on the out-of-test data to be analysed. Nevertheless, the performance of our trained CNN on our particle masks during the training-validation-test stage demonstrates that even simple neural network architectures can be promising for the 3D tracking of microswimmers with optical asymmetry, and further work in this direction might yield better outcomes than that achievable with traditional ML models. 

\end{document}